\documentclass[12pt]{iopart}
\usepackage[dvips]{graphicx}
\usepackage{amssymb}

\def\Z{{\Bbb{Z}}}
\def\R{{\Bbb{R}}}
\def\dd{{\rm d}}
\begin{document}
\frenchspacing
\def\picbox#1#2{\fbox{\vbox to#2{\hbox to#1{}}}}
\def\bra#1{\langle#1|}
\def\ket#1{|#1\rangle}
\def\ave#1{\langle #1 \rangle}
\def\parc#1#2{\frac{\partial #1}{\partial #2}}
\def\rot{\textrm{rot}}
\def\grad{\textrm{grad}}
\def\pa{\partial}
\def\scalar#1#2{\langle#1|#2\rangle}
\def\diag{{\rm diag}}
\def\card{{\rm card}}

\def\asin{{\rm arcsin}}
\def\acos{{\rm arccos}}

\def \grf#1#2#3{\vbox{\hbox to#1{\hfil \footnotesize #2 \hfil}\vskip-5mm
    \hbox to#1{\includegraphics[angle=-90, clip=, width=#1]{#3}}}}


\title{Uni-directional transport properties of a serpent billiard}

\author{Martin Horvat and Toma\v z Prosen}

\address{Physics Department, Faculty of Mathematics and Physics, 
University of Ljubljana, Slovenia}

\eads{\mailto{martin@fiz.uni-lj.si}, \mailto{prosen@fiz.uni-lj.si}}  
\begin{abstract}
We present a dynamical analysis of a {\it classical billiard chain --- a channel} with parallel semi-circular walls, which can serve as a model for a bended optical fiber. An interesting feature of this model is the fact that the phase space separates into two disjoint invariant components corresponding to the left and right uni-directional motions. 
Dynamics is decomposed into the {\it jump map} --- a Poincare map between the two ends of a basic cell, and the {\it time function} --- traveling time across a basic cell of a point on a surface of section.
The jump map has a mixed phase space where the relative sizes of the regular and chaotic components depend on the width of the channel. For a suitable value of this parameter we can have almost fully chaotic phase space. We have studied numerically the Lyapunov exponents, time auto-correlation functions and diffusion of particles along the chain. As a result of a singularity of the time function we obtain marginally-normal diffusion after we subtract the average drift. The last result is also supported by some analytical arguments. 
\end{abstract}
\submitto{\JPA}
%
\section{Introduction: uni-directional billiard channels}
The discussion of classical and quantum dynamics of spatially extended
billiard chains, either with periodicity or disorder, is a promising field of research
with a variety of direct applications, e.g. in nanophysics, fiber optics, 
electromagnetic cavities, etc. It is fair to say that studies in spatially extended 
billiard systems have been
under represented as compared to a vast amount of work which has been dedicated to
billiards on bounded domains. Nevertheless, one has to mention several basic results
in this type of systems. First, one can study the escape rates from finite portions
of an infinite billiard chain, like the Lorentz channel \cite{gaspard1}.
Second, one can study classical transport properties, such as diffusion and
transport of heat along the billiard chains in order to understand the dynamical
(microscopic) origin of the macroscopic transport laws \cite{alonso1,alonso,casati}.
Third, one can study the relation between deterministic diffusion in
a classical billiard chain, Anderson-like dynamical localization in the 
corresponding quantum chain, and the nature
of its spectral fluctuations \cite{dittrich1,dittrich2,dittrich3}.
And fourth, there is an interesting effect of 
localization transition in the presence of correlated disorder, which has been
studied in the case of billiard chains both theoretically \cite{izrailev} and
experimentally \cite{kuhl}.
\par
\begin{figure}[!htb]
\begin{center}
  \hbox{
     \begin{minipage}[b]{8cm}
     {\footnotesize(a)}\\* \vskip-1cm {\centering
     \includegraphics[width=7.5cm]{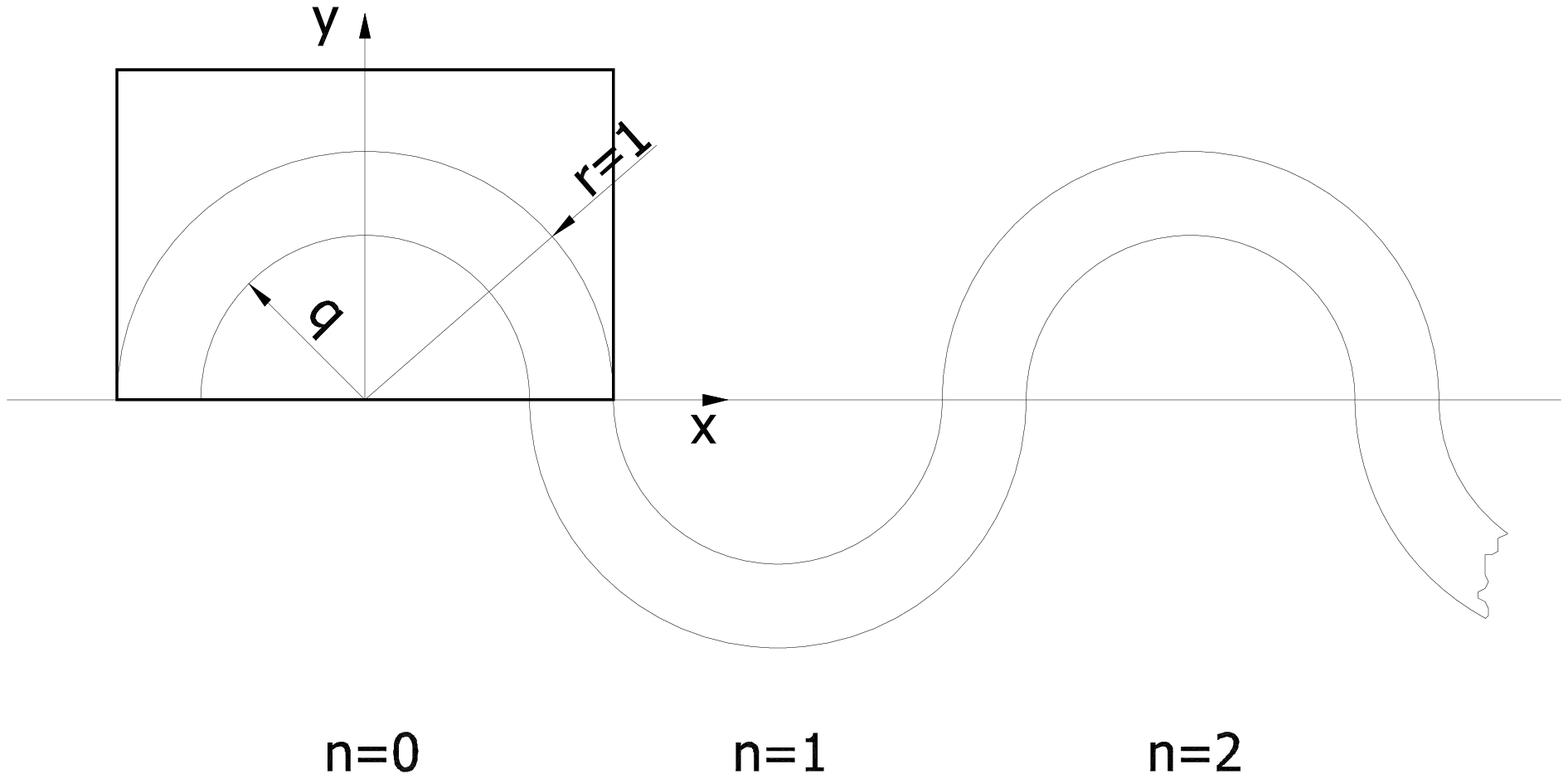}}
     \par\vspace{0pt}
   \end{minipage}  
   \begin{minipage}[b]{8cm}
     {\footnotesize (b)}\\*\vskip-1cm{\centering
     \includegraphics[angle=-90, width=7.5cm]{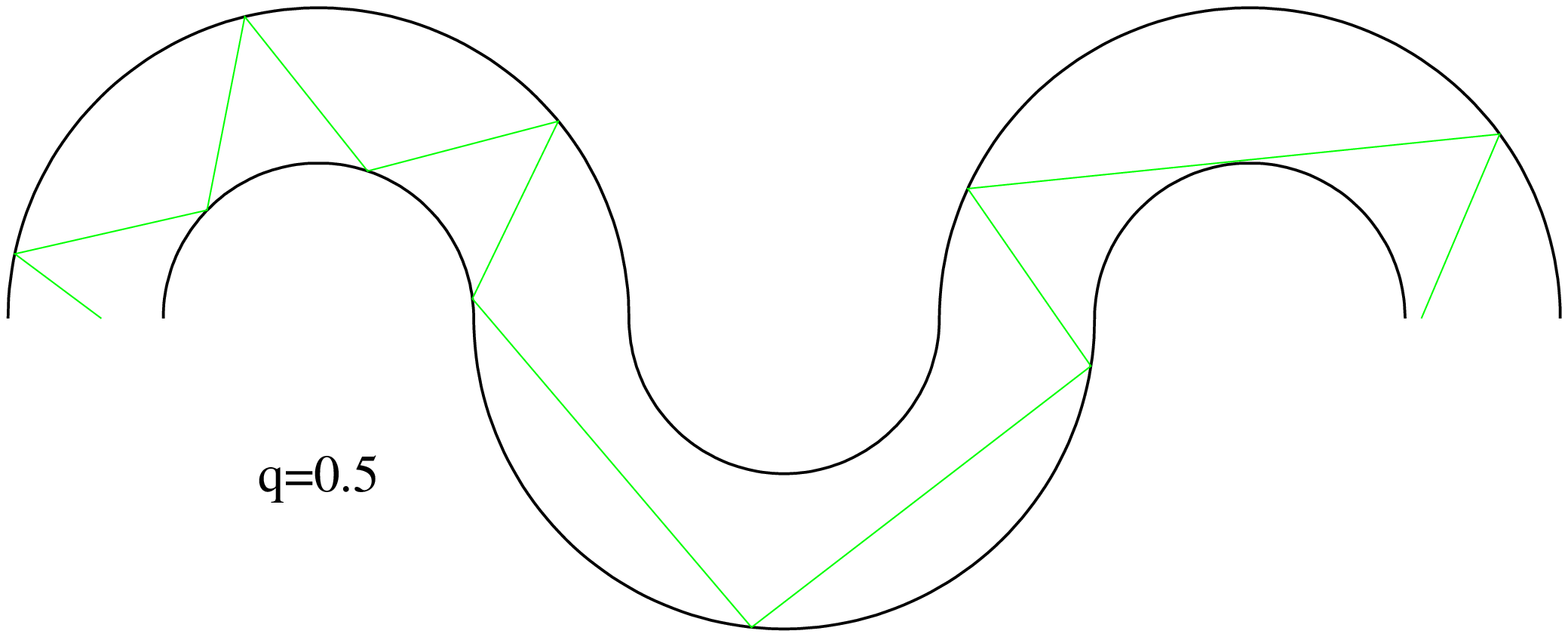}}
     \par\vspace{0pt}
   \end{minipage}
  }  %
\end{center}
\label{pic:shema}
\caption{A schematic picture of the serpent billiard model (a) with one basic
cell put in a rectangular frame, and index $n$ labeling the consecutive cells. 
A typical trajectory is indicated in (b).}
\end{figure}
In this paper we discuss a class of classical billiard channels with an unusual and distinct 
dynamical property, namely uni-directionality of the ray motion along the chain.
Specifically we focus our study on a particular billiard chain --- a channel with 
parallel semi-circular walls which we name as {\it serpent billiard}. 
The billiard under discussion is built as a periodic composition of semicircular rings of radii 
$q \in [0,1)$ and $1$, for the inner and outer circular arc, respectively. The geometry of this
billiard chain and illustration of the ray dynamics are shown in fig.~\ref{pic:shema}.
By construction the phase space separates into two disjoint invariant components corresponding 
to 
the left and right uni-directional motions, corresponding to two different signs of the
angular momentum as defined with respect to the origin (center) of the ring of the current 
billiard cell. Within each cell the angular momentum is conserved. Further, it is obvious that upon
the transition between one cell to another the sign of angular momentum, as calculated with 
respect two the centers of adjacent cells, remains unchanged.
Therefore, particles traveling from left to right initially will do so forever and will thus 
never be able to change the direction of travel, and similarly for the motion in the opposite 
direction, so that these two motions constitute two disjoint invariant halves of the
phase space. Nonetheless, as we shall show below, the dynamics inside each invariant half of the
phase space may be (practically) totally chaotic and ergodic.
\par
We note that this property of uni-directionality can be proven for a more general billiard channel 
which is bounded by an arbitrary pair of {\em parallel} smooth curves. Namely, it is easy to prove the following observation.
\\\\
Let the billiard motion in $\R^2$ be bounded by two smooth $C^1$ curves
${\cal C}_j$, $j=1,2$, with natural parameterizations $s\to \vec{r}_j(s)$.
The curves ${\cal C}_1$ and ${\cal C}_2$ should never intersect and they should be parallel in 
the following sense: for any $s\in\R$, a line ${\cal L}(s)$ intersecting ${\cal C}_1$ 
{\em perpendicularly} at $\vec{r}_1(s)$ should also intersect ${\cal C}_2$ 
{\em perpendicularly}, say at point $\vec{r}_2(\tau)$ defining a map $\tau=\sigma(s)$. 
The function $\sigma : \R \to \R$ 
should be monotonously increasing invertible function, i.e. $\sigma'(s) > 0$ for all $s$,
or in other words, the lines ${\cal L}(s)$ {\em should not intersect} each other inside the billiard region.

\begin{figure}[!htb]
  \begin{center}
  \includegraphics[width=7.5cm]{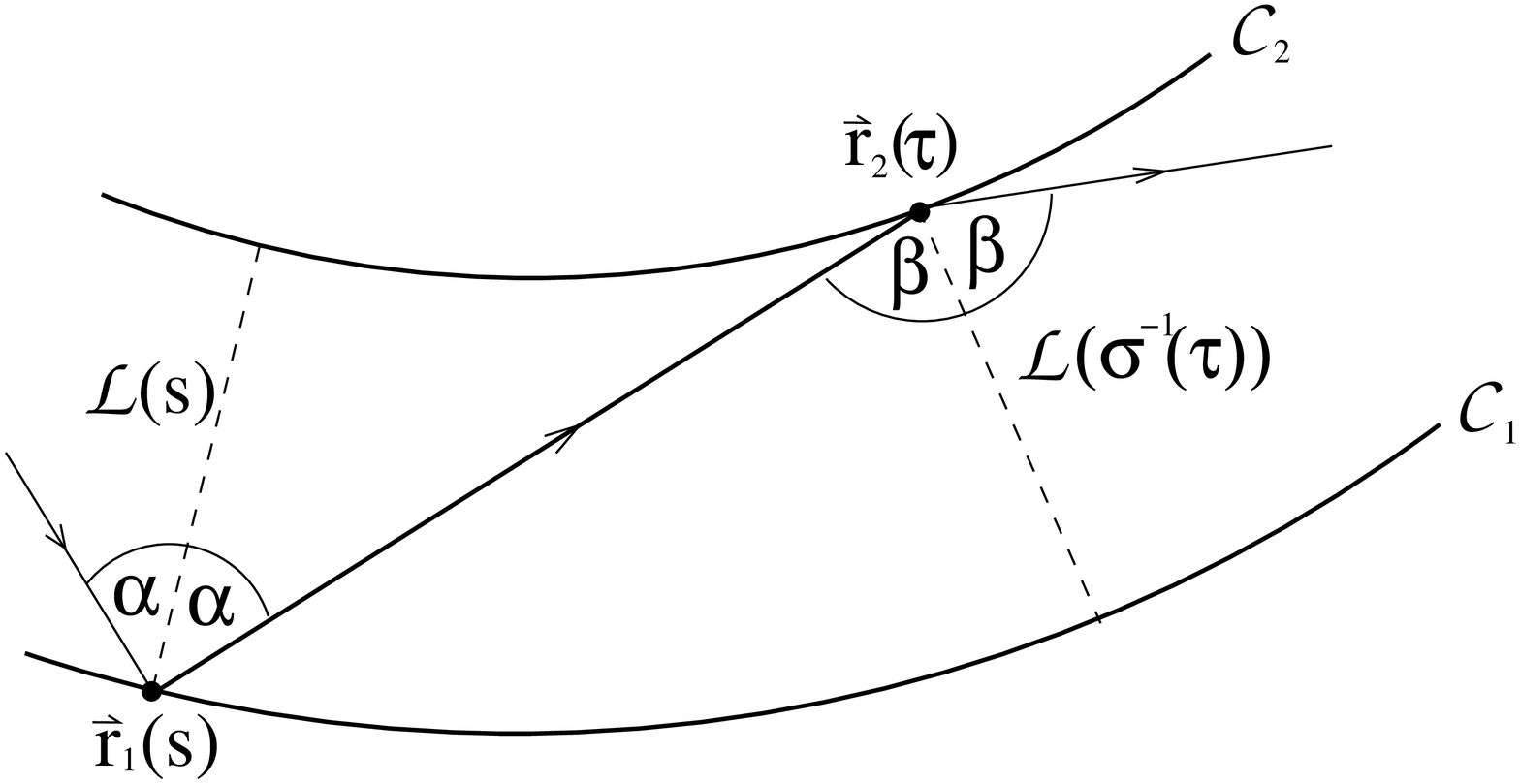}
  \includegraphics[width=7.5cm]{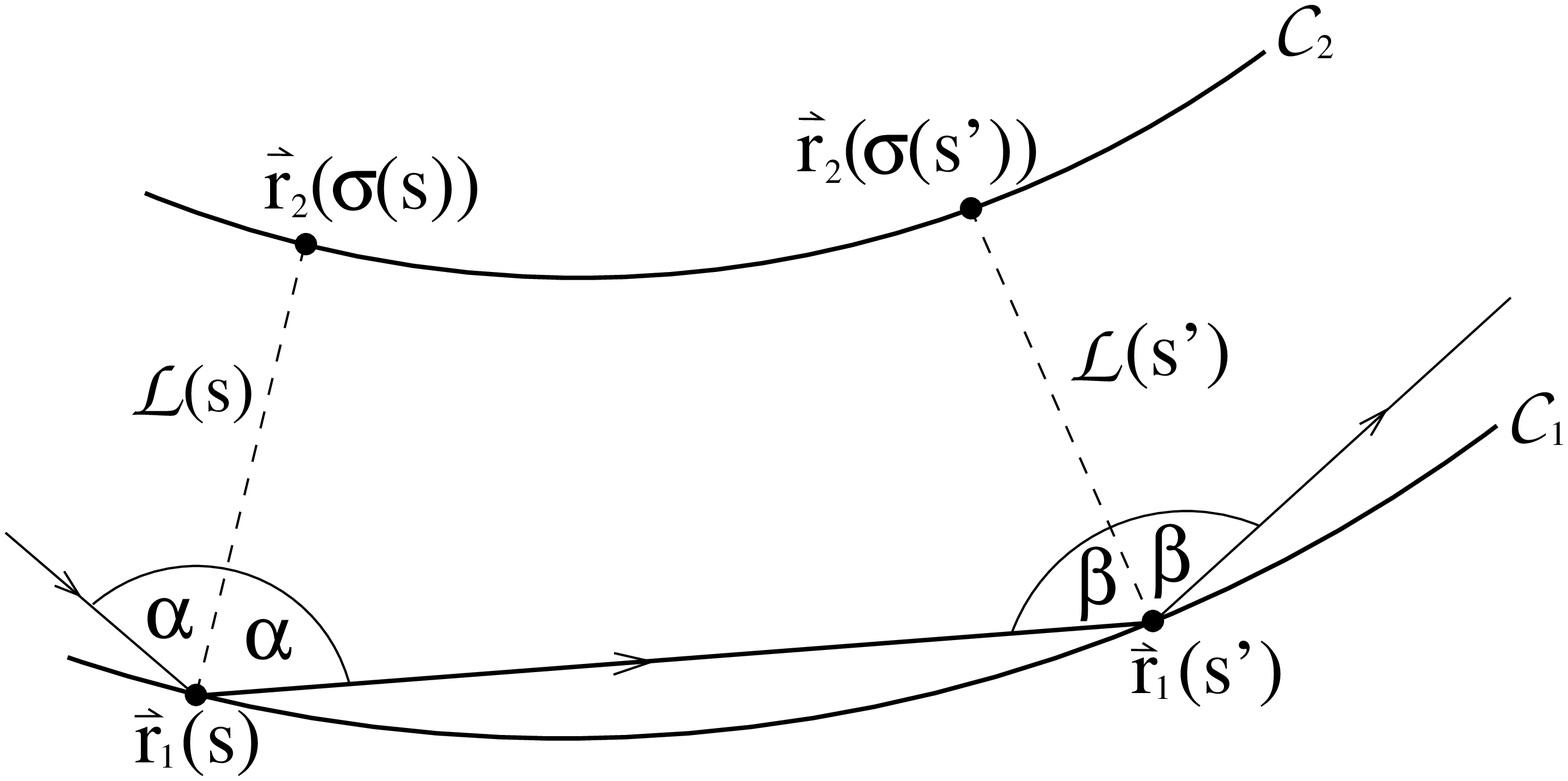}
  \end{center}
  \hfil(a)\hfil\hfil(b)\hfil
  \caption{The schematic diagram shows two possible cases (a,b) of subsequent collisions needed for the proof of unidirectionality.}
  \label{pic:diag}  
\end{figure}

Then the billiard motion is {\em uni-directional}, i.e. the sign of the 
tangential velocity component $\vec{v}\cdot (\dd/\dd s)\vec{r}_j(s)$ 
stays constant for all collision points of an arbitrary trajectory.
\\\\
To prove this observation, it is sufficient to consider two 
subsequent collisions of a segment of trajectory with a velocity of unit 
length $|\vec{v}|=1$. We may assume the first 
collision to take place at $\vec{r}_1(s) \in {\cal C}_1$ and write 
$\sin\alpha:=\vec{v}\cdot(\dd/\dd s)\vec{r}_1(s) > 0$. 
Then we consider two possible cases:\\
(a) The
next collision happens on the curve ${\cal C}_2$, say at the point $\vec{r}_2(\tau)$. Then the angle of
incidence again writes as $\sin\beta:=\vec{v}\cdot(\dd/\dd \tau)\vec{r}_2(\tau)$. 
$\alpha$ and $\beta$ are the angles between the segment of the trajectory and lines ${\cal L}(s)$ and
${\cal L}(\sigma^{-1}(\tau))$, respectively. Since the latter two do not cross inside the billiard region,
it follows that the sign of $\alpha$ and $\beta$ should be the same (positive). We may have $\beta=0$ only if $\alpha=0$, i.e. when the motion takes place along ${\cal L}(s)$ which is a periodic orbit.
See fig.~\ref{pic:diag}a.\\
(b) Another possibility is that the next collision happens with the same curve, i.e. at 
$\vec{r}_1(s')$. Writing $\sin\beta = \vec{v}\cdot(\dd/\dd s)\vec{r}_1(s')$ we again observe that 
the sign of the angles $\alpha$ and $\beta$ should be the same (positive), considering that the
other ends of the lines ${\cal L}(s)$ and ${\cal L}(s')$ at points $\vec{r}_2(\sigma(s))$
and $\vec{r}_2(\sigma(s'))$ should lie on the same side of the trajectory segment since the 
latter should not cross ${\cal C}_2$. 
See fig.~\ref{pic:diag}b. 
Thus we have proven that ${\cal L}(s)$ is a family of marginally stable periodic orbits, of vanishing 
overall measure, which separates the phase space of the billiard into two halves of unidirectional motions.

It is perhaps worth stressing that the conditions of parallelism as expressed in the statement imply also that
the width of the channel should be constant, $|\vec{r}_1(s)-\vec{r}_2(\sigma(s))| = {\rm const}$.
\\\\
We should note that detailed understanding of the dynamics of such a class of billiards may have
useful applications, in particular in fiber optics, electromagnetic waveguide propagation, etc.
In the following sections we shall concentrate on the dynamics of the specific serpent billiard model, 
which we shall analyze in terms of a special version of the Poincare map, 
the so-called jump map.
Then we shall describe analytically and numerically the average transport velocity, 
deterministic diffusion and correlation functions of the model.
\section{Dynamics of the serpent billiards}
Let us consider the Hamiltonian dynamics of a particle in a serpent billiard channel. 
We are considering a classical point particle with a fixed velocity of unit size. 
Due to uni-directionality of the motion, as shown above, we may freely choose to consider only
forward propagation --- in the positive direction of $x$ axis --- as it is shown in the example
of fig.~\ref{pic:shema}b. The forward dynamics of the billiard can be written in terms
of dynamics within a given basic cell and a transition to an adjacent basic cell.
In order to fully describe the dynamics we only need to know a Poincar\' e-like transformation
which maps coordinates of an entry into a cell to the coordinates of an exit and the time spent between entry and exit (the entry into an adjacent cell). Thus we formulate the dynamics of our 
billiard chain in terms of a jump map model \cite{zumofen}. 
We shall define the jump map more precisely below.
\par
Let the particle enter the basic cell on the left end at the initial position $x$ and with the
horizontal velocity component $v_x$ and travel to the right end where it exits at 
the position $x'$ and velocity $v'_x$. To clarify the notation we introduce a 
phase space {\em entry section} 
${\cal S}_{\rm L}$ and an {\em exit section} ${\cal S}_{\rm R}$
$$
 {\cal S}_{\rm L} = \{(x,v_x): x\in [-1,-q], v_x\in [-1,1]\},\quad 
 {\cal S}_{\rm R} = \{(x,v_x): x\in [q,1], v_x\in [-1,1]\}.
$$
The dynamical mapping of an entry point to an exit point shall be denoted by
${\bf G} : {\cal S}_{\rm L} \to {\cal S}_{\rm R}$,
\begin{equation}
  (x',v_x') = {\bf G}(x,v_x).
\label{eq:mapG}
\end{equation}
The map $\bf G$ can be expressed analytically since the billiard in the circular ring is integrable.
Since the lengthy expression is not very illuminating we give its
explicit form in the Appendix.
In order to be able to apply $\bf G$ again, we have to transform the current exit position 
$x'$ to the entry position of the next basic cell by a map 
${\bf S} : {\cal S}_{\rm R} \to {\cal S}_{\rm L}$ 
$$
  (x'',v_x'') = {\bf S}(x', v_x') = (x' - 1 - q, v_x').
$$
With this transformation the conservation of angular momentum around the center of the current cell
is broken which implies non-integrability of the model. We should mention that our serpent
billiard falls into the category of {\em semi-separable} systems \cite{Prosen96}.
The propagation of a particle from one basic cell to another can then be stated 
in terms of a single map ${\bf F}:{\cal S}_{\rm L} \to {\cal S}_{\rm L}$
\begin{equation}
  {\bf F} = {\bf S} \circ {\bf G}.
\end{equation}
We will refer to ${\bf F}$ as a {\it jump} or {\it Poincare map} and to the phase space 
${\cal S}_L$ as a {\it surface of section} (SOS). 
In terms of the map ${\bf F}$ the dynamics over the whole 
channel is decomposed into spatially equidistant snapshots. 
In order to maintain the whole physical information about the dynamics we have to 
introduce the {\em time function} $T:{\cal S}_{\rm L}\to\R^+$, 
i.e. $T({\bf x})$ measures the time needed by a particle to travel from 
the entry point ${\bf x}=(x,v_x)\in {\cal S}_{\rm L}$ to the other end of the basic cell. 
The pair $({\bf F},T)$ now represents the {\it jump model} corresponding to our billiard channel.
Again, the time function $T({\bf x})$ could be written out explicitly, though with a cumbersome
expression, so we put it in the Appendix. 
However, we should note that numerical routines for computing the map ${\bf F}$ and the function $T$ are very elementary and efficient.
 
\begin{figure}[!htb]
  \begin{center}
  \includegraphics[angle=-90, width=9cm]{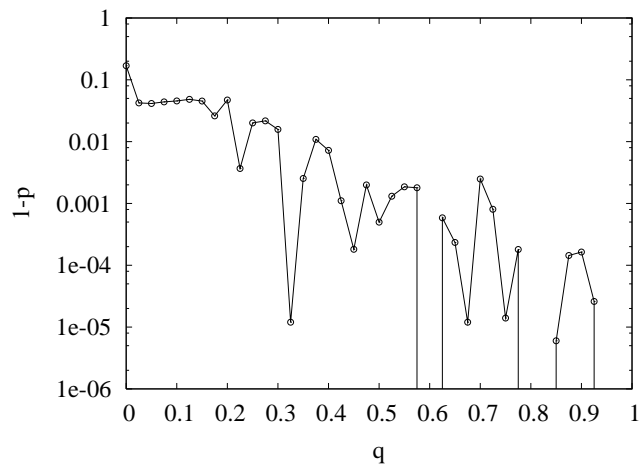}
  \end{center}
  \caption{
Numerical estimation of the total relative area of regular SOS components $1-p$, where $p$ is the relative area of the chaotic component of SOS, for the jump map is shown as a 
function of the parameter $q$. The ratio is obtained by sampling of 1000 random 
trajectories starting inside the chaotic component 
over a phase space grid of a size $1000\times 1000$. The length of trajectories in number of 
jumps was $10^6$.}  
 \label{pic:ratio}
\end{figure}

As a useful illustration of the gross dynamical features of the model we plot in fig.~\ref{pic:phase_space}
the phase portraits of the Poincar\' e-jump map ${\bf F}$ for different values of the parameter $q$.
We observe that the jump map has in general a mixed phase space with chaotic and regular regions coexisting
on SOS. We also observe that the phase portraits are symmetric in $x$ around the mean radius $(1+q)/2$.
We note that the chaotic component is always dominant in size over the regular components and for certain regions of parameter $q$ values the regular components are practically negligible so the jump map appears to become (almost) 
fully chaotic and ergodic. One nice example where we were unable to locate a
single regular island has the parameter value $q=0.6$. In fig.~\ref{pic:ratio} we plotted the relative area of regular SOS components as a function of parameter $q$. 
%
\def \grfa#1#2{%
	\vbox{%
	  \hbox to7cm{\hfill \footnotesize  #1 \hfill}\vskip-3mm
          \hbox to7cm{\includegraphics*[bb = 21 27 402 572, angle=-90, width=7cm]{#2}}}}
\begin{figure}[!htb]
\hskip1cm
\vbox{
 \hbox to14cm{\grfa{q=0}{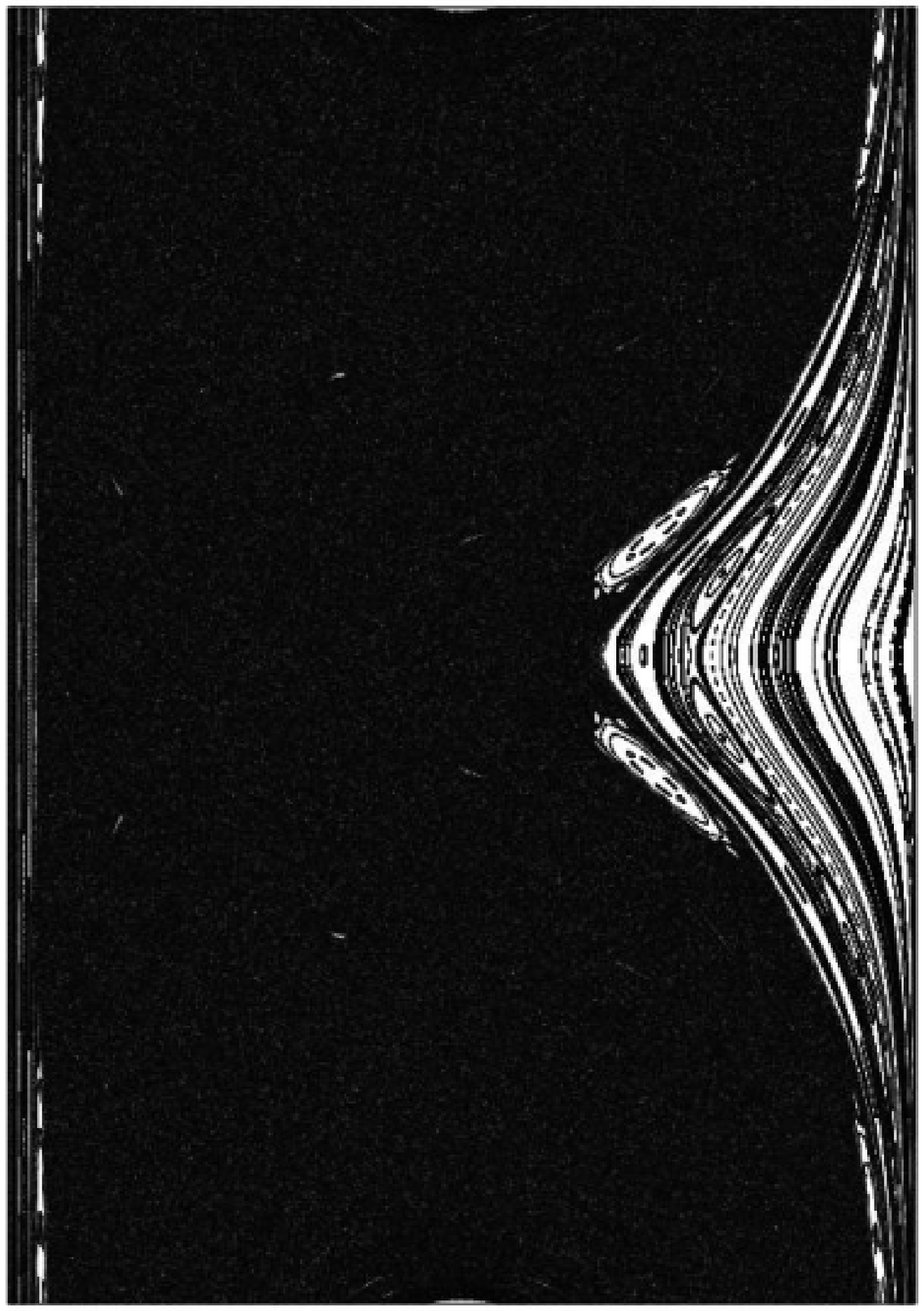}
  \grfa{q=0.3}{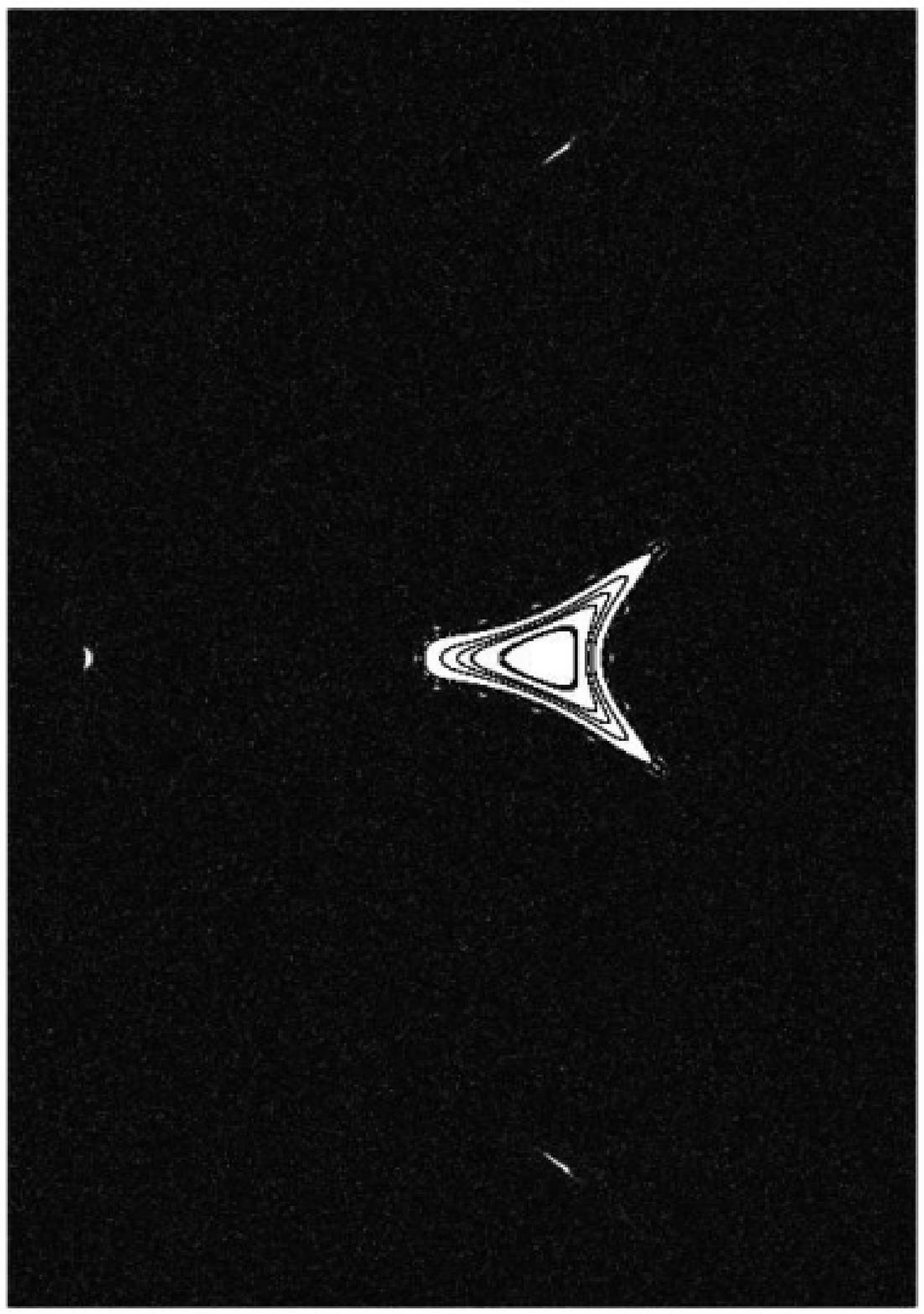}}
  \vskip1mm
  \hbox to14cm{\grfa{q=0.6}{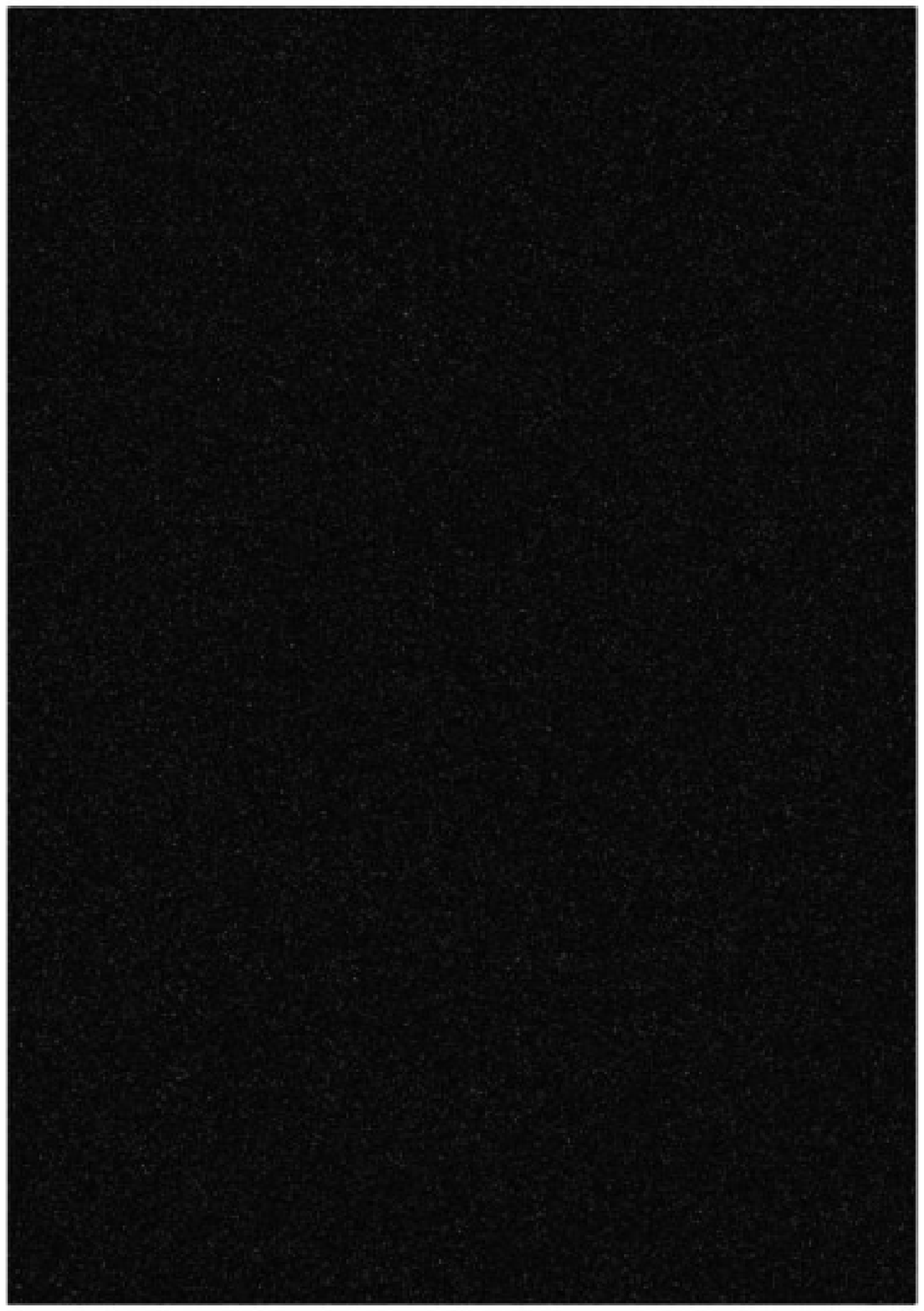}
  \grfa{q=0.9}{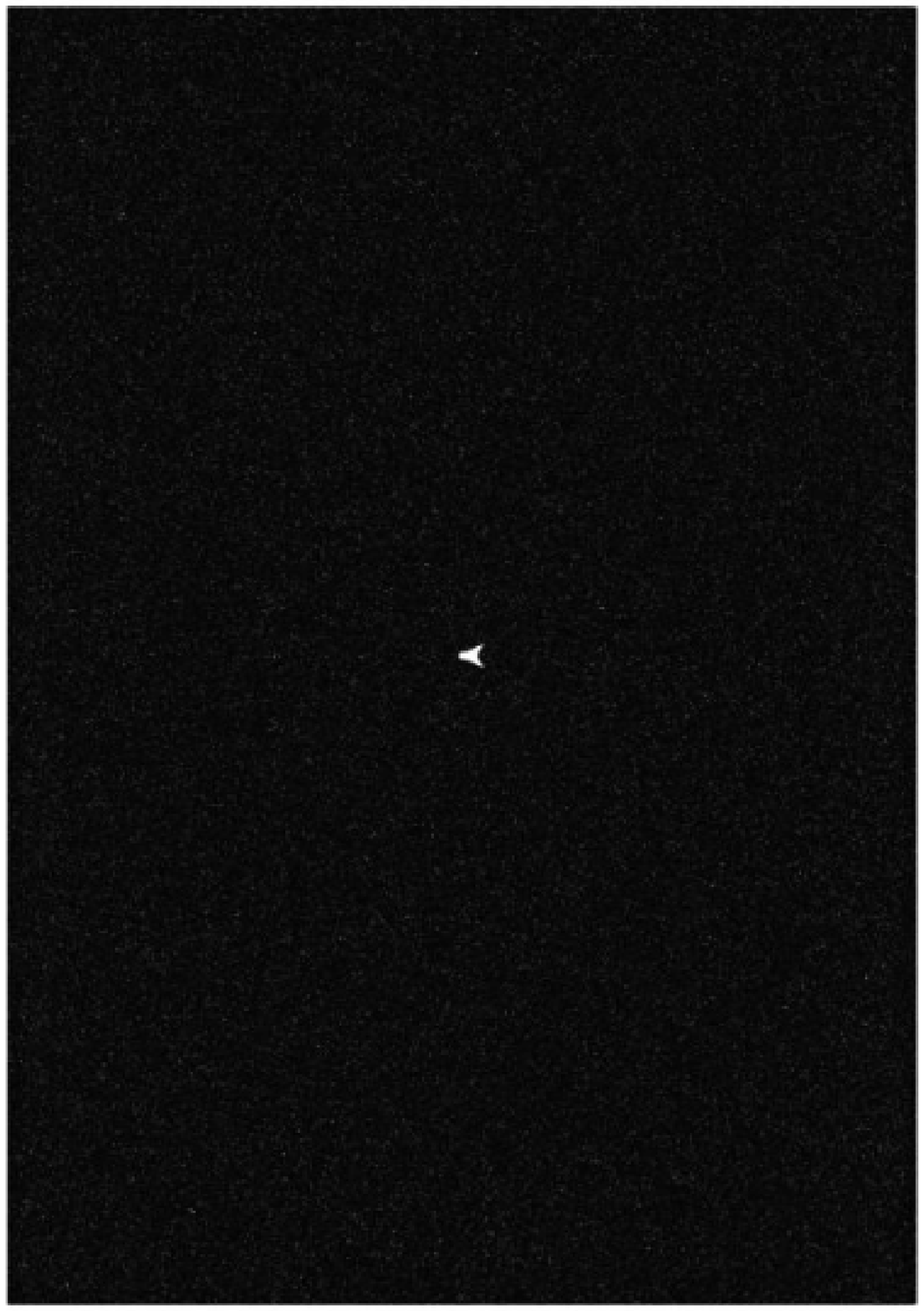}}
}
  \caption{Gallery of phase space portraits of the jump map ${\bf F}$ for different values 
of the parameter $q$. Horizontal axis: $x$, vertical axis: $v_x$. 
Each diagram shows $10^4$ successive iterations of 400 different 
initial points distributed randomly over SOS.} 
  \label{pic:phase_space} 
\end{figure}

\begin{figure}[!htb]
  \begin{center}
  \includegraphics[angle=-90, width=9cm]{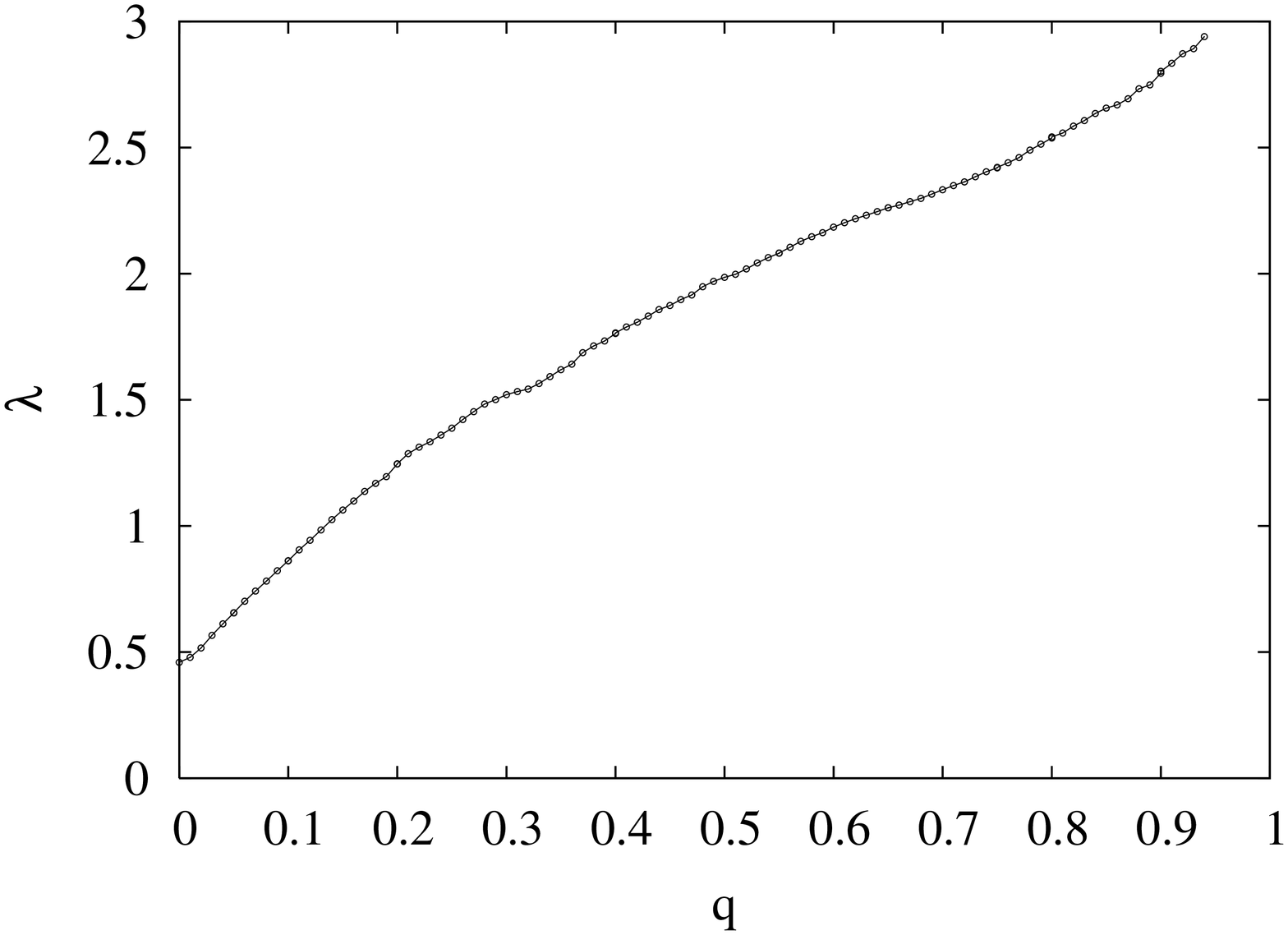}
  \end{center}
  \caption{Lyapunov exponent $\lambda$ of the jump map $\bf F$ 
averaged over the dominating chaotic component of SOS as a function of the
parameter $q$. The average was computed from 1000 trajectories of length $10^5$.}
  \label{pic:lyap} 
\end{figure}

%
In order to quantify the exponential instability of trajectories inside the chaotic component of SOS
we have measured the average Lyapunov exponent $\lambda$ (as described e.g. in \cite{reichl}).
The result for $\lambda$ as a function of $q$ is shown in fig.~\ref{pic:lyap}. 
We observe that the chaoticity as measured by $\lambda$, being equal to the dynamical Kolmogorov-Sinai 
entropy, is always positive and is increasing monotonically with $q$. This trend was somehow 
intuitively expected, as the number of collisions within the jump increases with $q$.
Namely, the bigger number of collisions between entry and exit sections implies less correlation between the
angular momenta of the adjacent cells meaning stronger integrability breaking.\par
\begin{figure}[!htb]
\begin{center}
  \includegraphics[angle=-90,width=7cm]{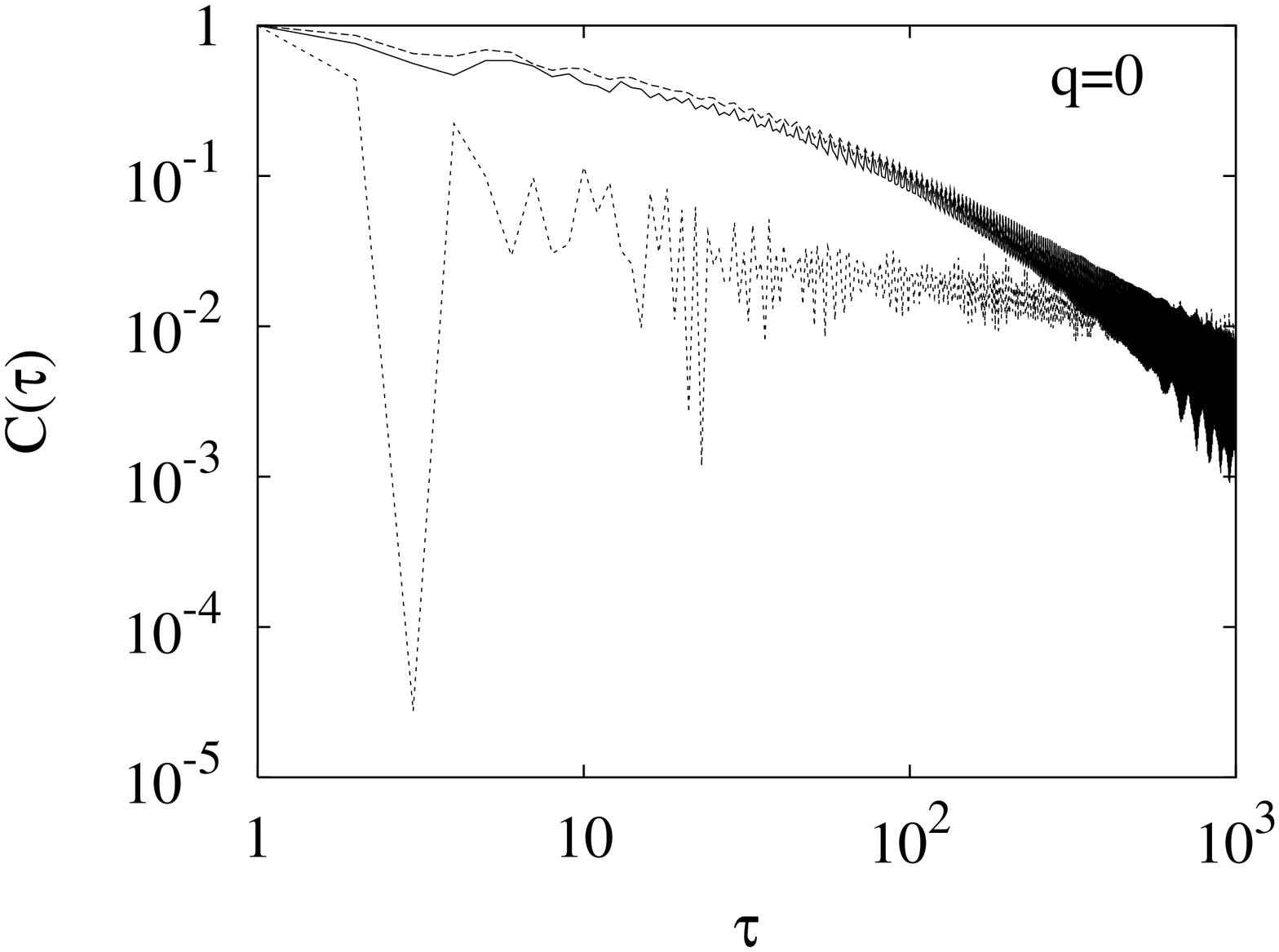}%
  \includegraphics[angle=-90,width=7cm]{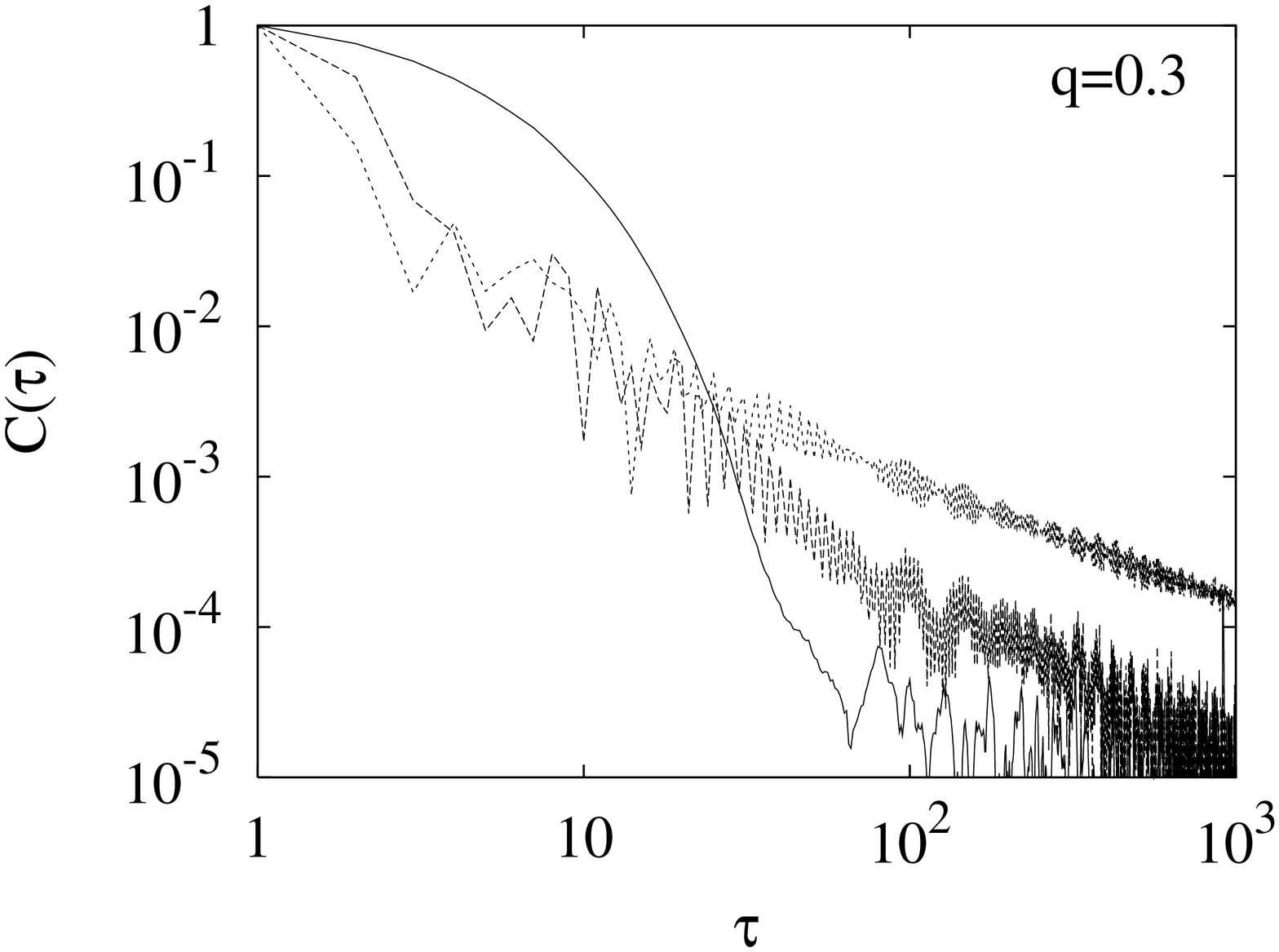}\\%
  \includegraphics[angle=-90,width=7cm]{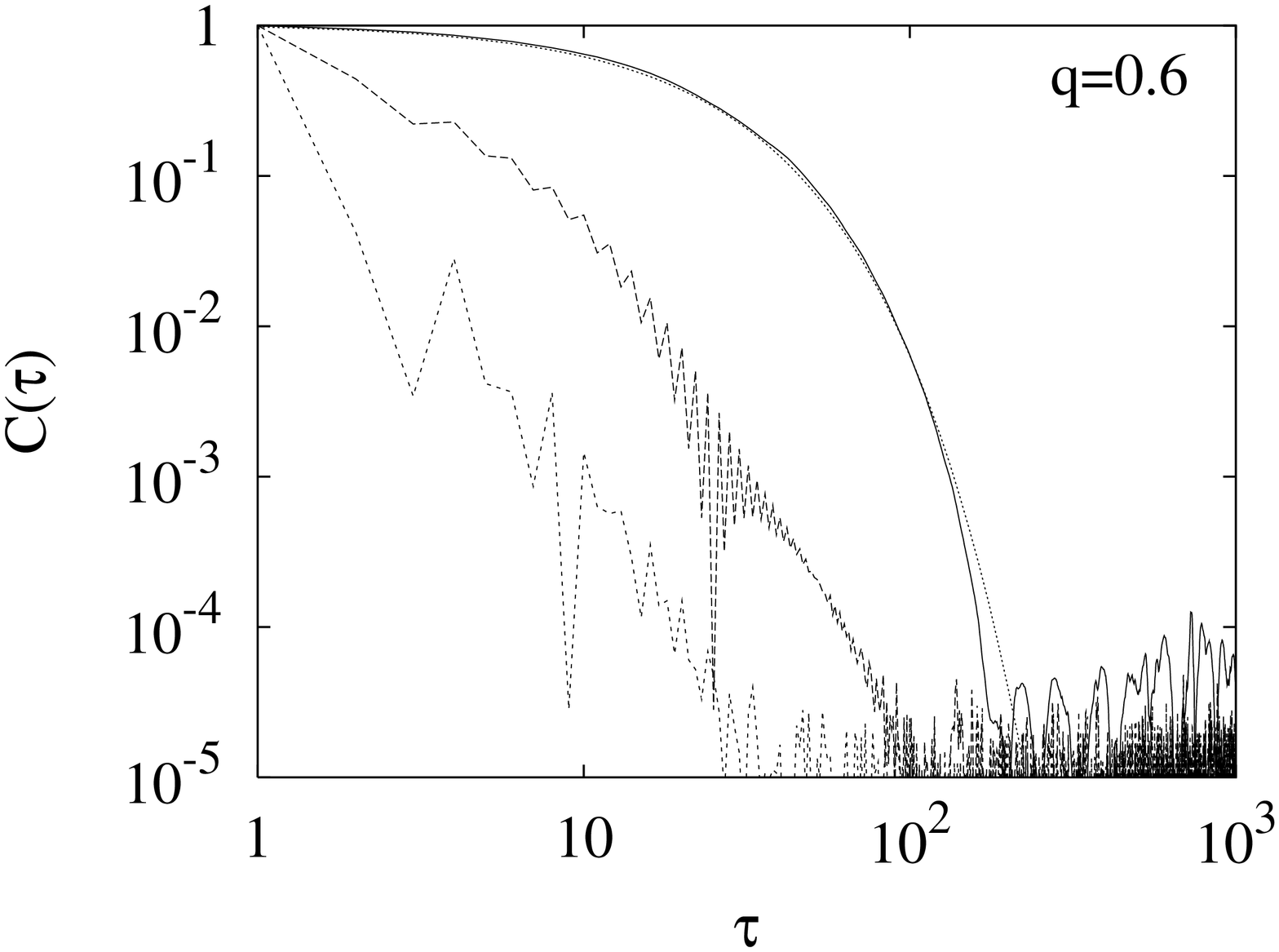}%
  \includegraphics[angle=-90,width=7cm]{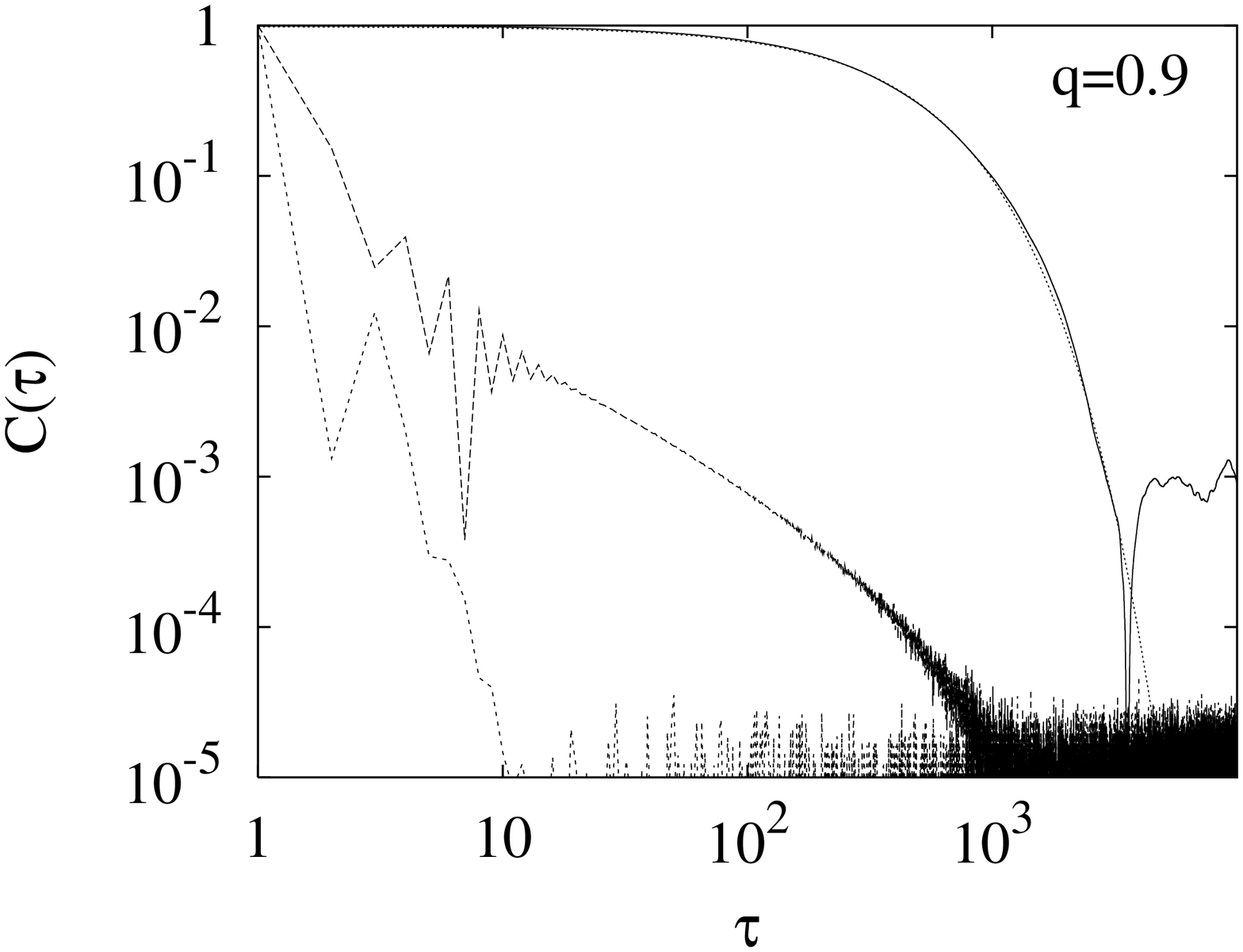}%
\end{center}
\caption{The average auto-correlation functions: 
$\ave{C_{T,T}({\bf x},\tau)}_{\Omega}$ (full curve), 
$\ave{C_{x,x}({\bf x},\tau)}_{\Omega}$ (dashed), and 
$\ave{C_{v_x,v_x}({\bf x},\tau)}_{\Omega}$ (dots) 
as functions of the number of jumps $\tau$. 
The calculation is performed using trajectories of length $5\cdot 10^{5}$ and averaged 
over $10^4$ initial conditions ${\bf x}$ over the chaotic SOS component. 
In the plot for $q = 0.6$ and $q=0.9$ we insert the exponentials
$\exp(-0.0508193\,\tau)$ and $\exp(-0.00234154\,\tau)$, respectively, in order ti guide the eye. The noisy plateaus indicate the level of statistical fluctuation.}
  \label{pic:corr} 
\end{figure}

%
Important fingerprints of dynamics, in many ways complementary to Lyapunov exponents, 
are the time-correlation functions. These reflect the mixing property of the system which implies decay
to equilibrium of an arbitrary initial phase space measure. Time correlation functions are also directly 
related to transport, which is studied in the next section, through a linear response formalism.
We discuss discrete time correlation function of the jump map between two observables $\phi({\bf x})$
and $\psi({\bf x})$, which is defined as
\begin{equation}
  C_{\phi,\psi}(\tau,{\bf x}) = \lim_{t\to\infty} 
\frac{ \sum_{k=0}^{t-1} \phi({\bf F}^{(k)}({\bf x})) \psi({\bf F}^{(k+\tau)}({\bf x})) }
     { \sum_{k=0}^{t-1} \phi({\bf F}^{(k)}({\bf x})) \psi({\bf F}^{(k)}({\bf x})) }.
\end{equation} 
The correlation functions are normalized such that always, $C_{\phi,\psi}(0,{\bf x})\equiv 1$,
even for observables which are not in $L^2({\cal S}_{\rm L})$.
In the sequel we consider autocorrelation functions of very regular observables such as 
the phase space coordinates, namely $C_{x,x}$ and $C_{v_x,v_x}$, and the autocorrelation function
of the time function $C_{T,T}$ which is even more interesting for two reasons: (i) $C_{T,T}$ is directly
related to particle transport as described in the next section, and (ii) $T$ is not
in $L^2({\cal S}_{\rm L})$ as discussed below.
Numerical data presented in fig.~\ref{pic:corr} strongly suggest that correlation function $C_{T,T}$ 
typically exhibits exponential decay $\sim\exp(-{\rm const}\tau)$ 
for most of values of $q$, except that for small parameter values 
$q < 0.3$ the initial exponential-like decay seems to turn into an
asymptotic power law decay $\sim t^{-{\rm const}}$.
On the other hand the correlation decay of non-singular observables, like $C_{x,x}$ and $C_{v_x,v_x}$
seems to behave as a power law for all values of $q$.
It is interesting to observe that the qualitative nature of correlation decay seems to be quite different for
different classes of observables such as $x$ compared to $T$. However, in all the cases time correlation
functions strongly decay which is a firm indication of the mixing property of the serpent billiard on the
chaotic component.
\par
\begin{figure}[!htb]
  \hbox{
     \begin{minipage}[b]{7cm}
     \centering
     \includegraphics[width=6cm]{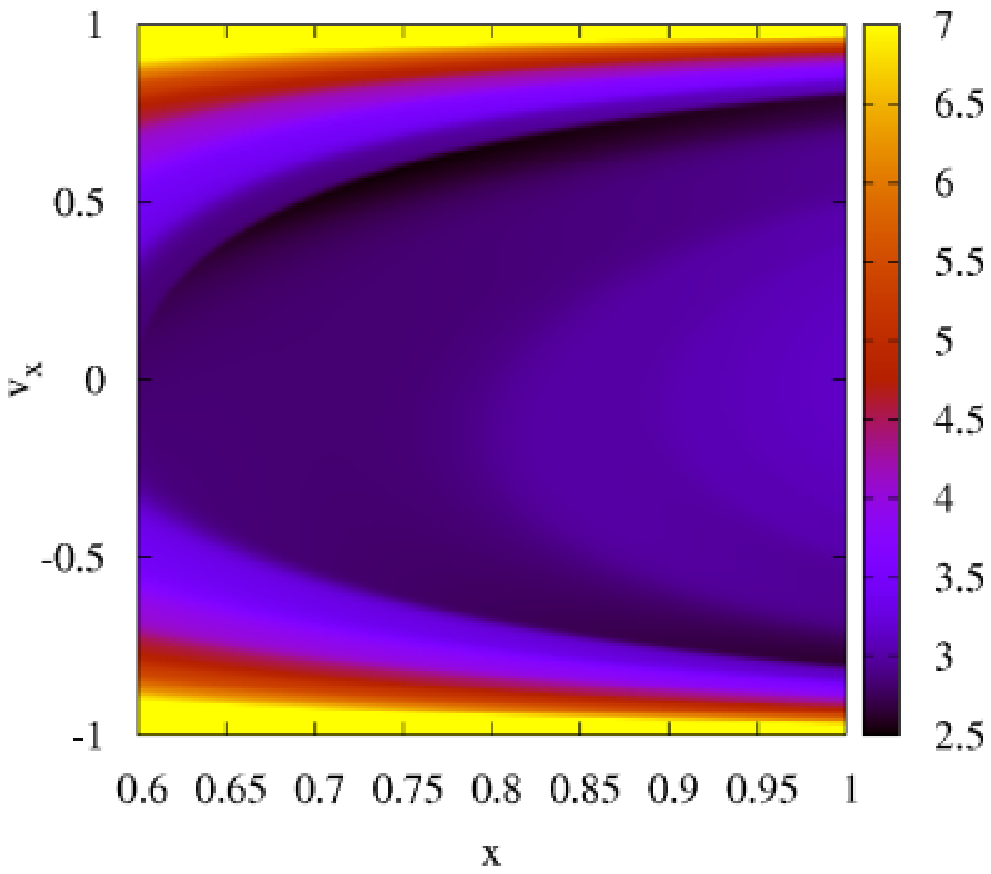}
     \par\vspace{0pt}
   \end{minipage}  
   \begin{minipage}[b]{8cm}
     \centering
     \includegraphics[angle=-90, width=8.5cm]{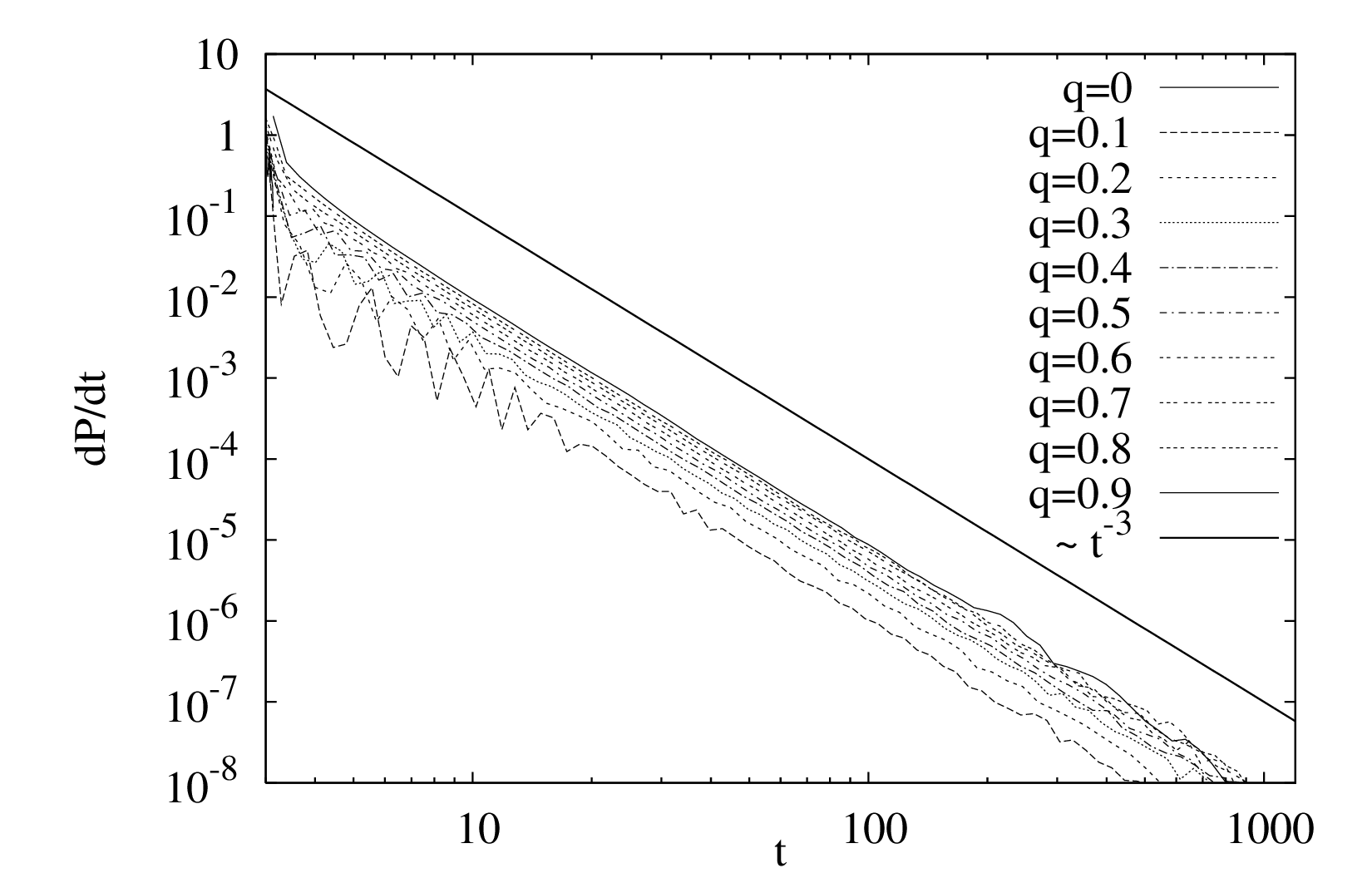}
     \par\vspace{0pt}
   \end{minipage}
  }
  \hfil (a) \hfil\hfil (b) \hfil
  \caption{(a) Density plot of the time function in logarithmic scale at $q=0.6$ and (b) the distribution $P(t)$ of the values of the time function for different $q$ indicated in the figure.
A thicker full straight line indicates $t^{-3}$ decay.}
  \label{pic:time_fun}  
\end{figure}

%
The time function $T({\bf x})$ is expected to have a singularity for $v_x \to \pm 1$ as it may take arbitrary long to traverse the semicircular ring with sufficiently small value of angular momentum. It is straightforward to show that this is a square-root singularity
\begin{equation}
  T(x,v_x) \sim (1-|v_x|)^{-1/2},\qquad |v_x| \sim 1.
\label{eq:sing}
\end{equation}
An example of the structure of the time function for $q=0.6$ is shown as a density plot in fig.~\ref{pic:time_fun}a. 
Another quantity which can illustrate the dynamical behaviour of observable $T$ is the
probability distribution $P(T)$ 
of times $T({\bf x})$ for a very long chaotic trajectory. Assuming that the system is ergodic on
the full SOS, the probability distribution $P(T)$ can be written in terms of the constant invariant measure on SOS
\begin{equation}
  P(t) = \int \delta (t - T({\bf x})) \dd^2 {\bf x}.
\end{equation}
Singularity (\ref{eq:sing}) implies the asymptotic form of the distribution 
\begin{equation}
P(t) \sim t^{-3}
\end{equation} 
for large $t$.
This asymptotic property does not essentially depend on the full ergodicity of the map, as we find the same asymptotic behaviour by numerical simulation of $P(t)$ for different values of $q$ as shown in fig.~\ref{pic:time_fun}b.
It is obvious that the only important condition for the universal decay of $P(t)$ is that the chaotic component should extend to the lines of singularity $v_x=\pm 1$.
The distribution $P(t)$ is very important, because it directly 
connects to the particle transport properties of the channel that 
are discussed in the next section.
\section{Transport properties}
%
%
Here we would like to examine the transport properties along our billiard chain in the context of the jump model. The basic cells are labeled 
with nonnegative 
integer $n\in \Z^+$ starting with $n=0$ and counting forward to the right. 
Transported length is measured as the number of traversed basic cells. 
This means that our length of travel is an integer but we should keep in mind that the physical length of one basic cell in the direction along the chain is $1+q$.
\par
Let us prepare an ensemble of particles on the SOS ${\cal S}_{\rm L}$ of the zeroth basic cell, $n=0$. 
The phase space is generally mixed and let the dominating chaotic component be denoted by $\Omega\subset {\cal S}_{\rm L}$.
We observe that each invariant phase space component may have its own transport properties and to obtain a clear picture 
of transport we have to test each component separately. The transport on regular components (islands) of SOS is obviously
ballistic since the islands are transporting with a constant and sharply defined average velocity.
Thus we concentrate on the more nontrivial case of transport on the chaotic component, namely in the following we choose a uniform initial distribution of the particles over the leading chaotic component $\Omega$: $\mu(\dd^2{\bf x}) = \dd^2{\bf x}/{\rm vol}(\Omega)$.\par
%

\begin{figure}[!htb]
  \begin{center}
  \includegraphics[angle=-90, width=10cm]{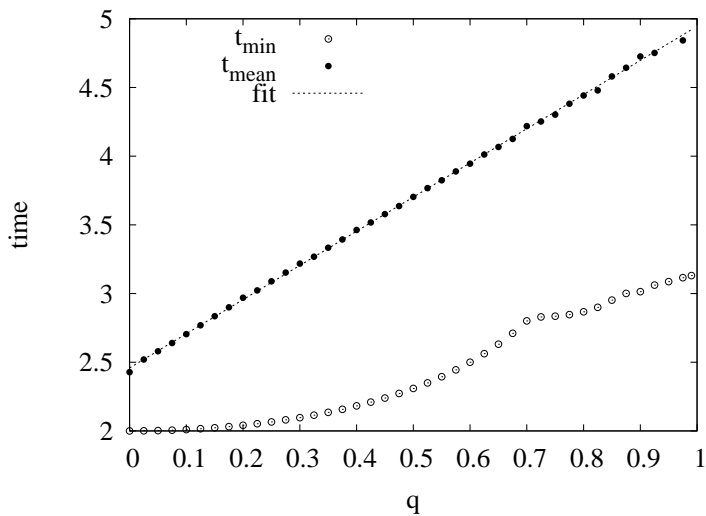}
  \end{center}
  \caption{Minimal time $t_{\rm min}$ and average time $t_{\rm mean}$ for traversing a basic cell over the chaotic
SOS component are shown as functions of parameter $q$.The curve that fits $t_{\rm mean}$ is the linear function $2.45863 +  2.48696q$.
The averaging and minimizing is performed on a trajectory of length $10^7$.}
  \label{pic:times}  
\end{figure}

\begin{figure}[!htb]
  \begin{center}
  \includegraphics[bb = 70 50 554 750, angle=-90,width=10cm]{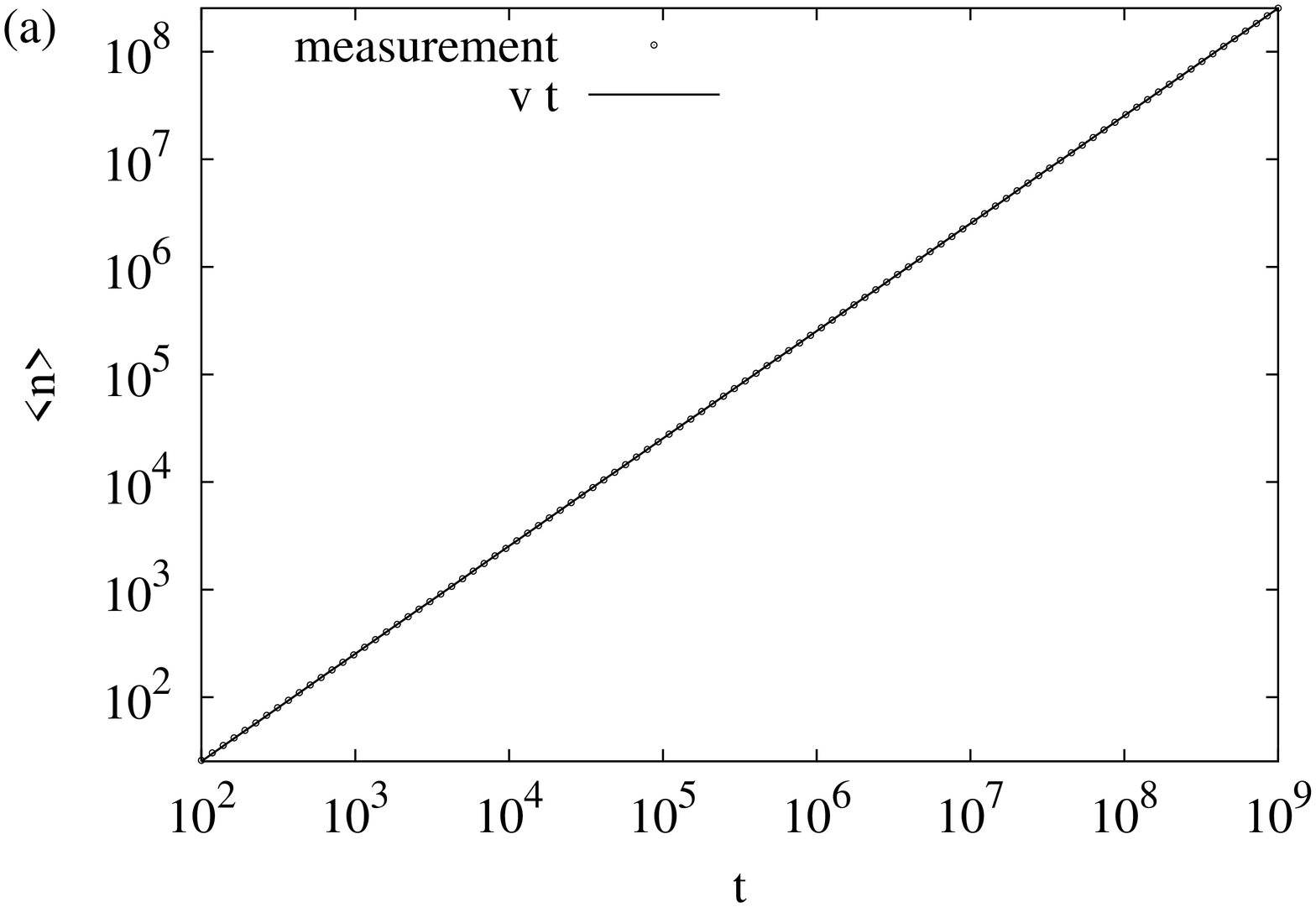}\\*
  \includegraphics[bb = 69 74 554 753, angle=-90,width=10cm]{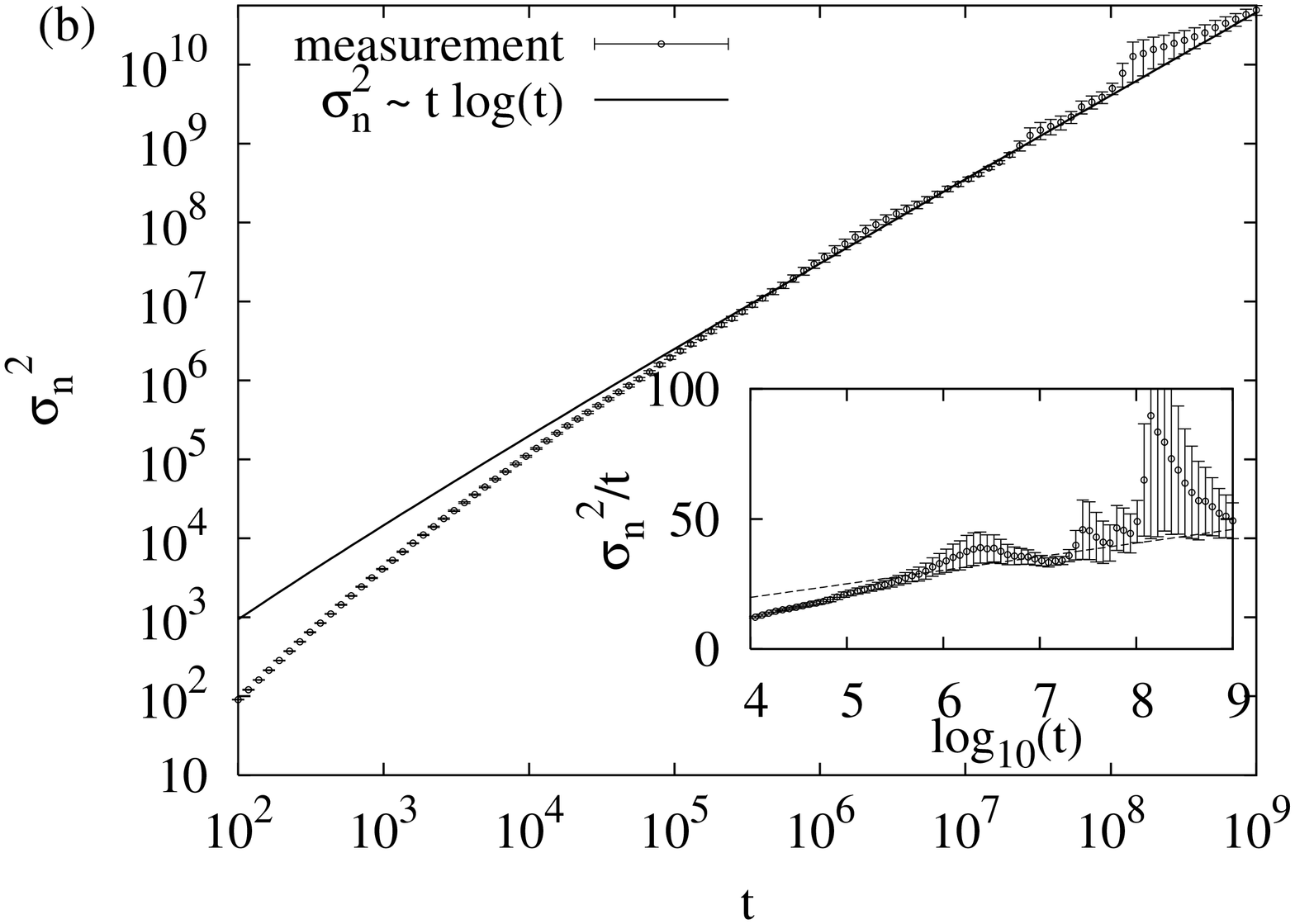}
  \end{center}
  \caption{The average position $\ave{n}_P$ (a) and the 
spread of particles $\sigma_n^2=\ave{n^2}_P -\ave{n}_P^2$ (b) 
as a function of time $t$ at $q=0.6$. The average transport velocity 
$v = 1/t_{\rm mean}$, where $t_{\rm mean}=3.94538$ is 
obtained from the data of fig.~\ref{pic:times}. 
The error bars represent the standard deviation of the 
measurement. The statistics is performed over an ensemble of
$25000$ initial points. The fitted line in the inset is given by $\sigma_n^2/t =-0.044374 + 5.2636\log_{10} t$}
  \label{pic:dif}  
\end{figure}

The transport of an ensemble of particles is described using the probability distribution $P(n,t)$ of particles over the basic cells $n\in \Z^+$ as a function of time $t$. Let us express this distribution in terms of a 
jump map ${\bf F}({\bf x})$ and the time function $T({\bf x})$. 
The time spent by a particle to traverse $m$ basic cells starting from the initial position ${\bf x}\in {\cal S}_L$ is calculated as
\begin{equation}\label{eq:timesum}
  T_m({\bf x}) = \sum_{k=0}^{m-1} T({\bf F}^{(k)}({\bf x})).
\end{equation}
The probability distribution of a single particle with initial coordinate ${\bf x}$ over the basic cells 
(labeled by $n\in \Z^+$) at time $t$ can be
written straightforwardly as
\begin{equation}\label{eq:pntx}
  P(n,t,{\bf x}) = \int_0^t \left\{
                   \delta(\tau-T_n({\bf x}))
                 - \delta(\tau-T_{n+1}({\bf x}))\right\} \dd \tau.
\end{equation}
By averaging $P(n,t,{\bf x})$ over an initial ensemble of particles, 
defined by 
\begin{equation}
\ave{f({\bf x})}_\Omega := \int_\Omega \mu(\dd^2{\bf x}) f({\bf x}),
\end{equation}
we obtain the distribution of particles over the cells
\begin{equation}\label{eq:pnt}
 P(n,t) = \ave{P(n,t,{\bf x})}_\Omega = \int_0^t \left\{p_n(\tau)-p_{n+1}(\tau)\right\}\dd\tau,
\end{equation}
where we are using the ensemble average distribution of times needed to traverse $n$ basic cells, denoted as $p_n(t)$,
\begin{equation}
 p_n(t) = \ave{\delta(\tau-T_n({\bf x}))}_\Omega.
\end{equation}
Let $\ave{\bullet}_P$ denote an average over the spatial distribution $P(n,t)$
\begin{equation}
 \ave{f(n)}_P = \sum_{n=0}^\infty f(n) P(n,t).
\end{equation}
%
Our aim is to obtain the time-asymptotic ($t\to\infty$) form of $P(n,t)$ where only contributions of far lying cells $n\gg 1$ are relevant. 
Since the auto-correlation functions in our model are decaying very fast, in particular the relevant $\ave{C_{T,T}({\bf x},t)}_\Omega$ (see the end of previous section), we may employ the central limit theorem and approximate the distribution of $T_n$, $p_n(t)$, in terms of distribution of $T$, $p_1(t)=\ave{\delta(t-T({\bf x}))}_\Omega$:
\begin{equation}\label{eq:approx_dpdt}
  p_n(t) = (\underbrace{p_1*\ldots*p_1}_{n})(t),\qquad n\gg 1.
\label{eq:conv}
\end{equation}
This approximation is very useful, because we can obtain the whole distribution $P(n,t)$ using a single function 
$p_1(t)=P(t)$ that can be easily measured and is already plotted in fig.~\ref{pic:time_fun}b. In the limit 
$n\to\infty$ we can treat the variable $T_n$ as infinitely divisible \cite{feller} and the parameter $n$ as a continuous 
variable. Then we can approximate the finite difference in $n$ in eq. (\ref{eq:pnt})
in terms of a derivative $\pa/\pa_n$ and calculate $P(n,t)$ as
\begin{equation}\label{eq:approx_pnt}
 P(n,t) \approx - \parc{}{n} \int_0^t  p_n(\tau) \dd \tau.
\end{equation}
Basic properties of the transport shall be described by the mean traversed length $\ave{n}_P$ and the spatial spread of the initial ensemble $\sigma_n^2 = \ave{n^2}_P - \ave{n}_P^2$. The mean $\ave{n}_P$ can be asymptotically, for $t\to\infty$, exactly expressed by the formula
\begin{equation}
 \ave{n}_P = \int_0^t \sum_{n=0}^\infty p_n(\tau)  \dd \tau.
\end{equation}
From the central limit theorem we immediately obtain the mean velocity as the inverse mean traverse time
\begin{equation}
 v = \lim_{t\to\infty} \frac{\ave{n}_P}{t}, \qquad  
 t_{\rm mean} = \frac{1}{v} = \int_0^\infty  t\,p_1(t) \dd t
\end{equation}
The average $t_{\rm mean}$, and the minimal time $t_{\rm min}$ to traverse a basic cell, 
as a function of $q$ are plotted in fig.~\ref{pic:times}.
The mean time $t_{\rm mean}$ is almost linearly increasing with increasing the parameter $q$. This reflects the obvious fact that the travel becomes
slower by narrowing the channel. The fact that the linear growth of $\ave{n}_P$ is indeed given by velocity $v$
is also demonstrated numerically for $q=0.6$ in fig.~\ref{pic:dif}a.
%
It is a little more tedious to obtain an analytical approximation for the average spreading width $\sigma^2_n(t) = \ave{n^2}_P-\ave{n}_P^2$. 

Essentially we need to control the growth of second moment of 
distribution $P(n,t)$. This distribution is given by eq.(\ref{eq:approx_pnt}) in terms of
 $p_n(t)$ which may be asymptotically expressed as a convolution of independent 
distributions $p_1(t)$ (\ref{eq:conv})
assuming sufficienly fast decay of temporal correlations as established
numerically. 
As a consequence, $p_n(t)$ inherits cubic singularity of $p_1(t)$, namely 
$p_n(t) \propto t^{-3}$.
This heuristic argument would suggest 
marginally normal diffusion $\sigma^2_n(t) \propto t \log t$, not essentially connected
to the strength of correlation decay but simply as a consequence of singularity of the 
time function. 

This observation can be formalized with a brief calculation. We stress that 
for large times $t\to\infty$ only cells with labels $n\sim v t \gg 1$ contribute appreciably to the probability distribution $P(n,t)$. In this regime we approximate $p_n(t)$ using (\ref{eq:approx_dpdt}), hence we can easily write its Fourier transform as
\begin{equation}
 \hat p_n(k) = \left ( \hat p_1(k) \right)^n = 
\exp\left(n\log \hat p_1(k)\right),\quad 
 \hat f(k) = \int_{-\infty}^\infty \exp({\rm i} k t) f(t)  \dd t.
\label{eq:pnk}
\end{equation}
The asymptotics of $\hat p_1(k)$ for $k\to 0$ will determine the asymptotics for $p_n(t)$ for long times $t\to\infty$. Thus we write the local expansion of $\hat p_1(k)$ around $k = 0$ explicitly taking into account the known asymptotics in time domain, $p_1(t\to\infty)\sim t^{-3}$, namely
\begin{equation}
  \log \hat p_1(k) = {\rm i} t_{\rm mean} k + 
                   \sigma_0^2 k^2(\alpha + \log k) +
		   O(k^3),\qquad k > 0,
\label{eq:p1k}
\end{equation}
where $\sigma_0^2$ and $\alpha$ are positive constants depending on the details of $p_1(t)$. Using eqs. (\ref{eq:p1k},\ref{eq:pnk}) and applying the inverse Fourier transform we find that the limiting distribution, namely $p_n$ for large $n$, is given by the formula
\begin{equation}\label{eq:limit_pn}
  p_n(t) = \frac{1}{\sigma_t}g\left(\frac{t-n t_{\rm mean}}{\sigma_t}\right), 
\qquad \sigma_t^2 = \sigma^2_0 n \log n, 
  \quad {\rm for}\quad n\gg 1.
\end{equation}
where $g(x) = \frac{1}{\sqrt{2\pi}}\exp(-x^2/2)$ is the standard Gaussian function. By inserting the above expression (\ref{eq:limit_pn}) into approximation of $P(n,t)$ (\ref{eq:approx_pnt}) and expanding in terms of parameter $n - vt$ we get the result
\begin{equation}
  P(n,t) = \frac{1}{\sigma_n}g\left(\frac{n-vt}{\sigma_n}\right),\qquad 
  \sigma^2_n  = \sigma_0^2 v^3 t \log(vt),\quad {\rm for}\quad
  n\gg 1.
\end{equation}
From this analysis we predict that the diffusion of initial ensemble of particles will distribute over the basic
cells with dispersion growing as $t\log t$. Thus we have shown analytically that our serpent billiard exhibits
marginally normal diffusion with a drift.
\par
%
We have tested our analytical results by performing extensive numerical simulations. An example of of $\sigma_n^2(t)$ for
$q=0.6$ is shown in fig.~\ref{pic:dif}b. We stress that our numerical data are indeed 
consistent with the marginally normal
diffusion.
\section{Summary and discussion}
In this paper we have analyzed a simple billiard chain, the so-called serpent billiard, 
with a special dynamical property of strictly uni-directional 
classical motion. We have also proven unidirectionality of motion for a more general class 
of billiard channels
with parallel walls.
The dynamics along the serpent billiard channel is described in terms of a variant of a jump model \cite{zumofen}, 
namely the jump-Poincar\' e map between the surfaces of section of two adjacent basic cells of the billiard, 
and the time function i.e. the time needed to traverse the basic cell as a function of the
position on the surface of section.
We have shown that the jump map is chaotic with generally mixed phase space, 
where the relative size of the largest chaotic component is generally increasing with decreasing of the
channel's width. The latter dependence is not strictly monotonic, because of bifurcations of regular components, 
but for narrow channels the chaotic component is typically largely dominant. 
For a considerable range of the parameter (denoting the channel's width) the jump map is even 
practically fully chaotic, ergodic, as no detectable islands of regular motion have been found. 
This does not mean that the
islands of stability can not exist for typical values of the parameter. We 
only wish to stress that it is easy to find parameter values 
(such as the case $q=0.6$ studied in the paper) 
for which all possible islands of stability are 
undetectably small for numerical (experimental) purposes.
Numerically measured maximal Lyapunov exponent shows that the chaoticity on a 
chaotic component is monotonically increasing with narrowing the channel.
\par
The transport of particles along the channel measured in the number of 
traversed basic cells exhibits a marginally-normal diffusion $\sigma_n^2\sim t\log t$ (when the drift term
is subtracted) due to square-root singularity of the time function. This singularity is a consequence of parallel walls, 
or saying in dynamical terms, it is due to a family of marginally stable bouncing ball trajectories
bouncing perpendicularly between the walls.\par
Besides its interesting and rather exotic dynamical properties, the model and its generalizations may also
be relevant for real world problems of transport, such as in optical fibers or 
wave-guides. These results may open even more 
interesting questions on the properties of quantum or wave transport of classically unidirectional billiard channels.
This is a subject of a subsequent publication \cite{mhtp}.

\section*{Acknowledgments}
Useful discussions with M. \v Znidari\v c, M. \v Copi\v c, F. Leyvraz, T. H. Seligman and G. Veble,
as well as the financial support by the Ministry of Education, Science and Sport of Slovenia are gratefully acknowledged.
\appendix
\section*{Appendix: Explicit jump map and time function}
Let the particle enter the basic cell at point ${\bf x} = (x,\cos\phi)\in {\cal S}_L$ end exit
at point  ${\bf x}' = (x',\cos\phi')\in{\cal S}_R$. Here we explicitly write the map ${\bf G}: {\bf x} \to {\bf x}'$, eq. (\ref{eq:mapG}), and the time function $T({\bf x})$. 
Due to conservation of the angulur momentum within a fixed cell we have the relation
$$
\Gamma = -x\sin\phi = x'\sin\phi'
$$
so we just have to give an explicit formula for the map $g:\phi\to\phi'$ and the time function 
$t(\phi)$. Then the remaining relation to specify the full map $G$ simply reads
$x' = -x \sin\phi/\sin g(\phi)$.
Let us write some auxilary variables
\begin{eqnarray*} 
\beta &=& \asin\Gamma,\quad \gamma = \asin(\Gamma/q),\\
\alpha' &=& \alpha + n\delta,\quad 
n =  \left \lfloor \frac{\pi - \alpha}{\delta}\right\rfloor,
\end{eqnarray*}
where $\lfloor x\rfloor$ is the largest integer not larger than $x$,
and $\alpha$ and $\delta$ are determined for each case separately below.
\\\\
If $\Gamma > q$ the particle is only hitting the outside wall.
Then, writing 
$\alpha = \pi - \phi - \beta$, $\delta = \pi - 2\beta$, we have
\begin{eqnarray*}
  \phi' = (n+1) \delta - \phi,\\
  t = \sqrt{1 + x^2  + 2x\cos\alpha}+
      \sqrt{1 + {x'}^2 + 2x'\cos\alpha'}+
      2n\cos\beta
\end{eqnarray*}
In the opposite case where $\Gamma < q$, writing $\delta = \gamma - \beta$ and 
$\Delta t = \sqrt{1 + q^2 - 2q\cos\delta}$, we have to discuss two cases:
(i) when the particle first hits the inner wall, $\phi < \pi/2$:
\begin{eqnarray*}
  \phi' = \left\{
    \begin{array}{ll}
      (n+1) \delta - \phi & n \;\textrm{odd}\\
      2\gamma - \pi + n \delta -\phi & n \;\textrm{even}
    \end{array}\right.\\
   t  = \sqrt{q^2 + x^2  + 2xq\cos\alpha} + n\Delta t +
        \left\{\begin{array}{ll}
         \sqrt{1 + {x'}^2 + 2x'\cos\alpha'}& n \;\textrm{odd}\\
         \sqrt{q^2 + {x'}^2 + 2x'q\cos\alpha'}& n \;\textrm{even}
       \end{array}\right.,  
\end{eqnarray*}
and (ii) when the particle first hits the outer wall, $\phi \ge \pi/2$:
\begin{eqnarray*}
\phi' = \left\{
    \begin{array}{ll}
      (n+1) \delta - \phi & n \;\textrm{odd}\\
      \pi - 2\beta + n \delta -\phi& n \;\textrm{even}
    \end{array}\right.\\
  t  = \sqrt{1 + x^2  + 2x\cos\alpha} + n\Delta t+
       \left\{\begin{array}{ll}
         \sqrt{q^2 + {x'}^2 + 2x'q\cos\alpha'}& n \;\textrm{odd}\\             
	 \sqrt{1 + {x'}^2 + 2x'\cos\alpha'}& n \;\textrm{even}
       \end{array}\right. .
\end{eqnarray*}
where $\alpha$ is equal to $\gamma - \phi$ in case (i) and $\pi - \phi - \beta$ in case (ii).
\section*{References}

\begin{thebibliography}{1}
%
\bibitem{gaspard1}Gaspard JP 1993 What is the role of chaotic scattering in irreversible processes? {\it Chaos} {\bf 3} 427-442
%
%
\bibitem{alonso1} Alonso D, Artuso R, Casati G and Guarneri I 1999 
Heat conductivity and dynamical instability, \PRL {\bf 82} 1859-1862

\bibitem{alonso} Alonso D, Ruiz A and de Vega I 2002 Polygonal billiards and transport: Diffusion and heat conduction, \PR E {\bf 66} 066131
%
\bibitem{casati}Li B, Casati G, Wang J 2003 Heat conductivity in linear mixing systems, \PR E {\bf 67} 021204
%
\bibitem{dittrich1} Dittrich T, Doron E, Smilansky U 1994 Classical diffusion, Anderson localization, and spectral statistics in billiard chains, \JPA {\bf 27} 79-114

\bibitem{dittrich2} Dittrich T, Mehlig B, Schanz H, Smilansky U 1997 Universal spectral properties of spatially periodic quantum systems with chaotic classical dynamics {\it Chaos Solitons \& Fractals} {\bf 8} (7-8) 1205-1227

\bibitem{dittrich3} Dittrich T, Mehlig B, Schanz H, Smilansky U 1998 Signature of chaotic diffusion in band spectra \PR E {\bf 57} (1) 359-365

\bibitem{izrailev} Izrailev F M 2003 Onset of delocalization in quasi-one-dimensional
waveguides with correlated surface disorder \PR B {\bf 67} 113402

\bibitem{kuhl} Kuhl U, Izrailev F M, Krokhin A A, and St\" ockmann H-J 2000
Experimental observation of the mobility edge in a waveguide with correlated disorder
Appl. Phys. Lett. {\bf 77} 633-635
%
%

\bibitem{reichl}Reichl, L E 1992 {\it The transition to chaos : in conservative classical systems : quantum manifestations} (New York [etc] : Springer-Verlag) 
%
\bibitem{feller}Feller W 1966 {\it An introduction to probability theory and
its applications} (New York [etc]: J. Wiley \& Sons,
cop.)
%
\bibitem{levy1}Shlesinger M F, Zaslavsky G M, Frisch U 1994 {\it L\'evy Flights and Related  Topics in Physics, Proceedings of the International Workshop held at Nice, France, 27-30 June 1994} (Berlin [etc]: Spinger-Verlag)
%
\bibitem{levy2}P\c{e}kalski A, Sznajd-Weron K 1998 {\it Anomalous Diffusion - From Basics to Application, Proceedings of the XIth Max Born Symposium Held at L\c{a}dek Zdr\'oj, Poland, 20-27 May 1998} (Berlin [etc]: Spinger-Verlag)
%
\bibitem{metzler}Metzler R, Klafter J 2000 The random walk's guide to anomalous diffusion: A fractional dynamics approach, {\it Phys. Rep} {\bf 339} 1-77
%
\bibitem{zumofen} Zumofen G, Klafter J 1993 Scale-invariant motion in the intermittent chaotic systems, \PR E {\bf 47} 851-863 

\bibitem{Prosen96} Prosen T 1996 Quantum surface of section method: Eigenstates and unitary quantum Poincar\' e evolution, Physica D {\bf 91} 244-277
%
\bibitem{mhtp}Horvat M, Prosen T, {\it in preparation}
%
\end{thebibliography}
\end{document}